# Title: Temporally-decoherent and spatially-coherent vibrations in metal halide perovskite

**Authors:** Depei Zhang[1*+], Xiao Hu[1*], Tianran Chen[1++], Douglas L. Abernathy[2], Ryoichi Kajimoto[3], Mitsutaka Nakamura[3], Maiko Kofu[3], Benjamin J. Foley[4], Mina Yoon[5], Joshua J. Choi[4], and Seung-Hun Lee[1, **]

[1] Department of Physics, University of Virginia, Charlottesville, Virginia 22904, USA.

[2] Neutron Scattering Division, Oak Ridge National Laboratory, Oak Ridge, Tennessee 37831, USA.

[3] Materials and Life Science Division, J-PARC Center, Tokai, Ibaraki 319-1195, Japan.

[4] Department of Chemical Engineering, University of Virginia, Charlottesville, Virginia 22904, USA.

[5] Center for Nanophase Materials Sciences, Oak Ridge National Laboratory, Oak Ridge, Tennessee 37831, USA.

* D.Z. and X.H. made equal contributions to this work.

+ Now at the Neutron Scattering Division, Oak Ridge National Laboratory, Oak Ridge, Tennessee 37831, USA.

++ Now at the NIST Center for Neutron Research, National Institute of Standards and Technology, Gaithersburg, Maryland 20899, USA.

** Corresponding Author. Email: shlee@virginia.edu

# Abstract

**The long carrier lifetime and defect tolerance in metal halide perovskites (MHPs) are major contributors to the superb performance of MHP optoelectronic devices. Large polarons were reported to be responsible for the long carrier lifetime. Yet microscopic mechanisms of the large polaron formation including the so-called phonon melting, are still under debate. Here, time-of-flight (TOF) inelastic neutron scattering (INS) experiments and first-principles density-functional theory (DFT) calculations were employed to investigate the lattice vibrations (or phonon dynamics) in methylammonium lead iodide (MAPbI$_3$), a prototypical example of MHPs. Our findings are that optical phonons lose temporal coherence gradually with increasing temperature which vanishes at the orthorhombic-to-tetragonal structural phase transition. Surprisingly, however, we found that the spatial coherence is still retained throughout the decoherence process. We argue that the temporally decoherent and spatially coherent vibrations contribute to the formation of large polarons in this metal halide perovskite.**

Metal halide perovskites (MHPs) have achieved striking success as low-cost photovoltaic and light-emitting devices [1-9]. Previous studies on MHPs suggest that their high performance in optoelectronic devices arises from the long carrier lifetimes, long carrier diffusion lengths, and exceptional carrier protection from defects [10-15]. Relevant underlying microscopic processes include polaron formation [16-23], exciton formation [24-30], electron-phonon coupling [25, 31-36], and phonon melting [10, 37]. For the three-dimensional (3D) hybrid organic-inorganic MHPs, it has been experimentally shown that the reorientation of the polarized molecules can assist the polaron formation, and thus prolong the charge carrier lifetime [21]. This was supported by a recent theoretical study based on the tight-binding model and first-principles density-functional theory calculations that reported increase of the polaron binding energy from 12 meV to 55 meV when molecular dynamic disorder is considered in addition to the electron-phonon coupling [19]. However, the purely inorganic MHPs without organic molecules can also achieve a moderate photovoltaic performance [38, 39], which indicates that other mechanisms that are directly related to the inorganic perovskite framework must also play indispensable roles in the optoelectronic properties of MHPs through their interactions with the charge carriers. The two relevant principal mechanisms of the inorganic framework for the photovoltaic properties are electron-phonon coupling [25, 31-36] and phonon melting [10, 37].

Electron-phonon coupling, i.e., the interaction between the charge carriers and the lattice vibrations, has been shown experimentally to affect various optical and electrical properties of MHPs. For example, a thermally induced blue shift of the bandgap was attributed to the population of a 1 THz optical Pb-I-Pb bending mode in MAPbI$_3$ [40]. Also, the thermally induced changes in the line shape, line width and intensity of the photoluminescence spectrum in various two-dimensional (2D) lead iodide perovskites were attributed to electron-phonon coupling [36]. Previous studies indicate that at room temperature the electron-phonon couplings in MHPs occur through the longitudinal optical (LO) phonons, such as the Pb-I-Pb bending or Pb-I stretching modes [25, 31, 40]. Zhu *et al*. argued that such electron-phonon coupling in 3D MHPs leads to the formation of large polarons which then protect the charge carriers from the defects and impurities [10, 17].

The so-called phonon melting describes a phenomenon that the phonon peaks that are well-defined in energy at low temperatures broaden upon heating and eventually become featureless continuum at high temperatures even though the system remains crystalline. The phonon melting, or more strictly, time-decoherent phonons, in MHPs [37, 41-46], are due to the intrinsic softness of their ionic crystals. Even when the phonons are decoherent in time, the system remains crystalline, i.e., the crystal-liquid duality, which leads to the 'band-like' charge carrier dynamics and glass-like phonons [10]. It was also argued that the glass-like phonon modes can further assist the optoelectronic performances in MHPs through their participation in the large polaron formation and hot carrier cooling processes [10]. Understanding the microscopic nature of the crystal-liquid duality and how it affects the optoelectronic properties of MHPs requires systematic studies of vibrational dynamics in MHPs as a function of temperature.

Here, we employ the time-of-flight (TOF) inelastic neutron scattering (INS) experiments and first-principles density-functional theory (DFT) calculations to investigate the phonon dynamics in a 3D MHP, MAPbI$_3$. Our study showed that optical phonons lose temporal coherence gradually with increasing temperature which nearly vanishes at the orthorhombic-to-tetragonal phase transition. During the temporal decoherence process, however, the spatial coherence is still retained. We argue that the liquid-like temporal decoherence and the crystalline spatial coherence of optical phonons yield both higher polarizability and spatial extension of localized phonon packet to form large polarons in this 3D lead halide perovskite.

The time-of-flight neutron scattering experiments were performed on an 8 g powder sample of CH$_3$NH$_3$PbI$_3$ to examine the vibrational motions. Fig. 1 (a) shows a color contour map of the neutron scattering cross section, $S(Q,\hbar\omega)$, as a function of momentum transfer, $Q$, and energy transfer, $\hbar\omega$, taken at 10 K. The phonon spectra are roughly composed of three regions: (i) the low energy part (0 ~ 10 meV), the mostly-inorganic modes, denoted as inorganic modes, which mostly involve the collective vibrations of inorganic atoms, including the Pb-I-Pb translation and Pb-I-Pb bending modes [47-49]; (ii) the intermediate energy part (10 ~ 30 meV), the organic-inorganic hybrid modes, denoted as hybrid modes, which are mixture modes of the inorganic motions, mainly the Pb-I-Pb rocking and Pb-I-Pb stretching, and the molecular rigid-body motions, i.e., CH$_3$NH$_3$ translation, spinning and libration; and (iii) the high energy part (30 ~ 400 meV), the nearly pure organic-atom vibrations, denoted as organic modes, which are dominated by the internal molecular motions, e.g., CH$_3$NH$_3$ twisting (or torsion), bending, rocking, stretching and deformation.

The identification of phonon modes for the experimentally measured peaks using DFT calculations and molecular dynamics (MD) has been challenging [50, 51]. We found that long-range van der Waals (vdW) interactions play a critical role in capturing the key features of experimental phonon spectra of MAPbI$_3$; we found that the non-local correction to DFT total energy using van der Waals density functional schemes, such as vdW-DF2 [52-55], needs to be included to reproduce the phonon spectra for MAPbI$_3$. For instance, the antisymmetric twisting mode of NH$_3$ and CH$_3$, $T_{NH_3-CH_3}$, illustrated in the inset of Fig. 2 (e), shifts to a lower energy by

~ 8 meV (the red-shaded curve on the bottom of Fig. 2 (e)) explaining the experimental data, compared to when vdW interactions are not included (the blue-shaded curve on the bottom of Fig. 2 (e)). Detailed information on the vibrational modes, including their energies and animations, and density of states as a function of energy, are shown on S.-H. Lee's website [56].

Our DFT calculations with the vdW-DF2 correction estimated the London dispersion interactions between the organic MA$^+$ molecule and its neighboring inorganic [PbI$_3$]$^-$ layer to be $E_{vdW}^{MA^+-[PbI_3]^-} \approx -0.912$ eV that is ~ 13.4 % of the total electronic energy between the two components. Thus, it is not surprising that the London dispersion interaction with a distance dependence of $\frac{1}{r^6}$ [57] has a significant influence on the restoring force, hence the phonon energy, of the antisymmetric twisting mode of the NH$_3$ and CH$_3$ group of the MA$^+$ molecule.

The inelastic neutron scattering intensities due to the vibrational motions were calculated for MAPbI$_3$, using the density functional theory calculations (see the SI Session F for details). As shown in Fig. 1(b), the powder-averaged calculated intensities reproduce well the measured ones shown in Fig. 1(a), on both the $Q$ and $\hbar\omega$ dependences. The experimental observation, shown in Fig. 1(a) and 2(e), that the peak centered at 11.7(1) meV is significantly broader than the one at 38.2(1) meV is also reproduced. As shown in the calculated phonon dispersion relations and vibrational energy fraction shown in Fig. 1 (c) and 1 (d), respectively, this difference is due to the fact that the 11.7 meV mode is a collection of hybrid modes that are dispersive along high symmetric directions in $Q$-space while the 38.2 meV mode is the lowest energy vibrational mode of the MA molecule internal motion and is very weakly dispersive even in the crystal. We stress that the $Q$ and $\hbar\omega$ dependences of $S(Q, \hbar\omega)$ measured at 10 K are well reproduced by the calculations and the dispersiveness of the hybrid modes at intermediate energies and of the inorganic modes at low energies is a clear evidence that the hybrid and inorganic modes are strongly collective, in other words, those vibrations are spatially and temporally coherent at 10 K.

Now let us see how the phonon spectrum changes with increasing temperature. Fig. 2 (a) – (c) show contour maps of the momentum-integrated neutron scattering intensity $S(T, \hbar\omega)$ as a function of temperature, $T$, and energy transfer, $\hbar\omega$ that show the temperature-evolution of the vibrational dynamics of MAPbI$_3$ over the entire energy range up to ~ 400 meV. As discussed before and shown as the black curves in Fig. 2 (d) – (f), at 10 K (orthorhombic phase ($Pnma$)), there exist temporally and spatially coherent vibrational modes over the entire energy range. However, as seen in Fig. 2 (d) – (f), the peaks below 50 meV become featureless in energy above the orthorhombic-to-tetragonal phase transition at ~ 165 K (see Fig. 2 (e), (f)). This indicates that the optical inorganic and hybrid vibrational motions lose coherence in time, i.e., those phonons are melted in the tetragonal ($I4/mcm$) and cubic ($Pm\bar{3}m$) phases above 165 K. In the orthorhombic phase below 165 K, the two types of the modes behave differently; the optical inorganic modes, denoted by $A_1$ (mainly the Pb-I-Pb bending modes), increase in strength as $T$ increases up to 100 K above which they weaken and melt as the system enters the tetragonal phase at 165 K (see Fig. 2 (f) and 2 (i)). The hybrid modes, denoted by $A_2$ (the mixed modes of rocking and stretching of Pb-I-Pb and translation and spinning of CH$_3$NH$_3$) and $A_3$ (the mixed modes of stretching of Pb-I-Pb and translation and libration of CH$_3$NH$_3$), on the other hand, gradually decrease in strength as $T$ increases and then melt in the tetragonal phase (see Fig. 2 (f) and 2 (i)). The organic modes, denoted by $A_6$ (CH$_3$-NH$_3$ twisting (or torsion) modes) in Fig. 2 (e) and 2 (h), and $A_7$ (symmetric and asymmetric stretching modes of CH$_3$ and NH$_3$) in Fig. 2 (d) and 2 (g), behave differently; upon warming the 38.2 meV mode, $A_6$, gradually weakens and melts at 155 K, similar as the hybrid modes do. On the other hand, the 400 meV mode, $A_7$, gradually broadens but survives all the way to the cubic phase. These different behaviors can be understood when one considers the charge distribution in the MA$^+$ molecule. In the MA$^+$ molecule the overall positive charge is mainly distributed in the NH$_3^+$ group and the CH$_3$ group is more or less neutral [21]. In the 38.2 meV mode, $A_6$, NH$_3^+$ twists twice more than CH$_3$ does (the amplitude ratio of NH$_3^+$: CH$_3$ = 2.2:1) as illustrated in the inset of Fig. 2 (e), and thus the mode gets affected as the motion of the neighboring negative I$^-$ ions melts upon warming. The higher energy organic modes such as $A_7$, on the other hand, involves the equal strength motion of CH$_3$ and NH$_3$, and thus get less affected by the melting of the inorganic and hybrid modes. See the detailed phonon animations on S.-H. Lee's website [56].

The observed energy continuum up to ~ 50 meV is an indication of the emergence of liquid-like vibrations upon heating, i.e., the temporally-decoherent phonons. To study the $T$-dependence of the liquid-like vibrations, we have chosen the valley regions at $6 \leq \hbar\omega \leq 8$ meV ($I_4$) and $20 \leq \hbar\omega \leq 22$ meV ($I_5$) where intensity is weak at 10 K (see Fig. 2 (e)). As shown in Fig. 2 (h), $I_4$ (blue squares) and $I_5$ (red diamonds) show a gradual increase as $T$ increases within the low-$T$ orthorhombic phase. One can ask if the energy continuum is due to the contribution from the incoherent rotational dynamics as each MA$^+$ molecule rotates more freely as $T$ increases. To quantitatively examine this possibility, we have fitted the very low energy $S(Q, \hbar\omega)$ to the previously reported point group theory analysis for the rotational dynamics of the MA$^+$ molecule [58]. As described in the SI Session C in details, the rotational contributions can be determined. The color-coded dashed lines in Fig. 2 (f) are the rotational contributions at different temperatures, clearly showing that the energy continuum up to 50 meV is indeed mainly due to the melting of the inorganic and hybrid vibrational dynamics rather than rotational dynamics. Similar measurements were performed on a partially deuterated MAPbI$_3$ sample (CD$_3$NH$_3$PbI$_3$), where the incoherent scattering from H atoms are suppressed, and the same liquid-like vibrations were observed at high temperatures. The lifetimes of different types of phonons were evaluated as described in the SI session D and E, and the results are shown in Fig. 3(a).

A question that naturally arises is what happens to the spatial coherence of the collective vibrations when the inorganic phonons and the hybrid phonons lose their coherence in time. To address this issue, we investigate the $Q$-dependences of the vibrational modes $S(Q)$ for MAPbI$_3$ as a function of temperature. Fig. 4 (a) – (c) show the color contour maps of $S(Q, \hbar\omega)$ taken with $E_i = 60$ meV, covering the inorganic and the hybrid modes, at three different temperatures, 10 K, 170 K and 350 K, respectively. At 10 K, $S(Q, \hbar\omega)$ exhibits well-separated peaks in $\hbar\omega$ that are broad and centered at around $Q \sim 5 - 7$ Å$^{-1}$. At 170 K and 350 K, $S(Q, \hbar\omega)$ exhibits an energy continuum, as discussed, due to the temporal decoherence. Surprisingly, however, the energy continuum exhibits similar $Q$-dependence as that of the temporally-coherent phonons observed at 10 K.

In order to quantitatively analyze the data, we integrated $S(Q, \hbar\omega)$ over three different $\hbar\omega$ regions to cover the inorganic and hybrid modes at the low energy and intermediate energy regions, respectively. The resulting $S(Q)$ is plotted in Fig. 4 (d) – (f). Upon heating, all the resulting $S(Q)$ exhibit a gradual shift to lower $Q$, which is due to the increasing vibrational amplitudes. In addition, for the low energy range of $2 \text{ meV} < \hbar\omega < 5 \text{ meV}$ shown in Fig. 4 (d), in the tetragonal phase ($T > 165$ K) $S(Q)$ exhibits an additional peak centered at lower $Q \sim 2 \text{ Å}^{-1}$ which is due to the rotational dynamics of MA$^+$ molecule and is well reproduced by the aforementioned group theoretical model for the rotational dynamics, $S_{rot}(Q)$, as shown as the color-coded dashed lines. For other higher energy regions, the rotational contribution is negligible. We have fitted the vibrational contribution $S_{vib}(Q) = S(Q) - S_{rot}(Q)$ to a phenomenological function for phonons $S_{model}(Q, T) = a + b\, S_{calc}(Q, 0 \text{ K})\, e^{-\Delta U_{iso} Q^2}$, with the pre-calculated phonon intensity at 0 K using OCLIMAX [59], $S_{calc}(Q, 0 \text{ K})$, and the constant background, $a$, the intensity amplitude, $b$, and in the Debye-Waller factor $e^{-\Delta U_{iso} Q^2}$, $\Delta U_{iso} = U_{iso}(T) - U_{iso}(0 \text{ K})$, with $U_{iso}(T)$ representing the $T$-dependent isotropic atomic displacement parameter, which is proportional to the square of the vibrational amplitude. As shown as the color-coded solid lines in Fig. 4 (d) – (f), $S_{model}(Q)$ can reproduce the experimental $S(Q)$ for all the three energy regions and all the temperatures. The fitted isotropic thermal factors, $\Delta U_{iso}$s, are shown in Fig. 3 (b). Here, we emphasize that the experimental $S(Q)$ cannot be explained if the optical phonons completely lose the spatial coherence. It is because in the case of spatially-decoherent vibrations of Pb and I atoms $S(Q)$ should exhibit a peak centered at a much higher $Q > 10$ Å$^{-1}$ as expected for the localized vibrations of individual atoms [60]. Instead, our observation that $S_{model}(Q)$ for the spatially coherent vibrations at 10 K weighted by the Debye-Waller factor can also account for the $Q$-dependence of vibrations at high $T$s means that the optical vibrations at high $T$s are coherent in space even though the vibrational amplitude increases with $T$. This tells us that when the optical vibrational modes start losing coherence due to the weak ionic bonds, that will be discussed later, upon heating, the de-coherence occurs first in time and the spatial coherence remains at high $T$s.

To understand the origin of the phonon melting behaviors observed in MAPbI3, we calculated the binding energy between nearest-neighboring atoms for three different systems, CH3NH3PbI3, another ionic crystal NaCl and a covalent crystal SiO2 for comparison. The binding energies between inorganic atoms in the two ionic crystals have a binding energy of ~ 3 - 4 eV that is much weaker than the bonding energy of ~ 10 eV of the covalent crystal SiO2. The intra-molecule bonding energies of the organic molecules of MHPs, such as the N-C and C-C bonding energies of ~ 9 eV and ~ 7 eV, respectively, are comparable to that of SiO2, which is expected because the intra-molecule bonds are covalent. Thus, the gradual phonon melting, i.e., the $T$-induced temporal-decoherence observed for the low energy vibrations in MAPbI3 is due to the weak ionic bonds between the inorganic atoms in MHPs. The survival of the time-coherence of the purely organic high energy vibrational modes at room temperature, on the other hand, is due to the strong covalent intra-molecule bonds. Furthermore, we have simulated the powder averaged phonon spectra of the three systems as a function of the temperature (see the SI Session F.3 for details). As shown in Fig. S6, the temperature dependent phonon spectra simulations show that the gradually decreasing phonon intensities, i.e., the phonon melting behavior, is a general feature for the ionic lattices with the weak ionic bonds. The phonon modes, in the covalent lattices, e.g., $SiO_2$, on the other hand, can survive at high temperatures. It should be noted that the rotational dynamics of the organic molecule may also enhance the phonon melting, as the $C_4$ mode of the $MA^+$ cation gets activated when the system enters the intermediate-temperature tetragonal phase from the low-temperature orthorhombic phase [58]. However, it is to be emphasized that the $MA^+$ $C_4$ rotational mode requires breaking of the multiple bonds between iodine and hydrogen atoms and thus its activation is a direct outcome of the weak ionic I-H bonds.

Note that $MAPbI_3$ is crystalline up to 650 K [61]. Thus, according to the Goldstone theorem the very low energy acoustic phonons below ~ 2 meV should exist as long as the system is crystalline. Indeed, a recent experimental study reported existence of temporally and spatially coherent acoustic phonons in $MAPbI_3$ at 350 K [62]. Our experimental results, on the other hand, showed that for the optical inorganic and hybrid phonons the temporal coherence is lost for $T >$ 165 K $\ll$ 650 K, which is consistent with previous Raman and neutron scattering studies on isostructural systems $MAPbX_3$ (X = Br [37, 41], Cl [42]). More strikingly, we have shown that those optical phonons retain spatial coherence. The observed spatial coherence of the optical phonons above 165 K may explain the previous results that phonon-mediated large polarons are formed in 3D MHPs, and the diameter of the polarons was theoretically determined for 3D MHPs to be ~ 50 Å [19] and was experimentally determined to be ~ 100 - 140 Å in terms of an exciton picture [63]. Previous optical measurement studies, such as Raman scattering, photoluminescence, and terahertz spectroscopy, showed that the electron-phonon couplings in MHPs occur through the optical phonons, such as the Pb-I-Pb bending and Pb-I stretching modes at room temperature [25, 31, 40]. For a phonon to mediate to form a polaron, the spatial coherence of the phonon should be larger than or at least comparable to the size of the polaron. These imply that the optical inorganic and hybrid phonons of MAPbI$_3$ must be spatially coherent at least over ~ 50 Å, which is consistent with our findings. Indeed, a theoretical study based on molecular dynamics simulations on anharmonic materials [64] showed that the temporal coherence and the spatial coherence of phonons can behave differently. Their argument is the following: the temporal coherence is associated with the phonon mean free path, i.e., the product of the phonon relaxation time and group velocity, reflecting the particle nature of phonons, which is consistent with the low thermal conductivities of MHPs [51, 65-69]. On the other hand, the spatial coherence represents the spatial extension of localized phonon packet reflecting the wave nature of the phonons. They showed that at some specific points in the Brillouin zone, the difference between the phonon mean free path and spatial extension can be giant, and even when the mean free path becomes very small, the spatial extension can be large, yielding a nonlocalized standing wave. To the best of our knowledge, our neutron scattering results are the first experimental observation of distinct temporal and spatial coherence of phonons.

The concept of the crystal-liquid duality has been actively studied in the family of solids called phonon-glass electron-crystals (PGECs), such as intermetallic clathrates, that exhibit high electrical conductivity but low thermal conductivity, and thus provide good candidates for efficient thermoelectric materials [10, 70]. Indeed, very recently the thermal conductivity of MHPs was found to be one of the lowest among all measured solid materials [71]. It would be interesting to see if the distinct temporal and spatial coherences of phonons found in MHPs are also a characteristic of PGECs that is involved in the microscopic mechanism of their thermoelectric properties.

# Acknowledgments

**Funding:** The work at the University of Virginia was supported by the U.S. Department of Energy, Office of Science, Office of Basic Energy Sciences under Award Number DE-SC0016144. A portion of this research used resources at the Spallation Neutron Source, a DOE Office of Science User Facility operated by the Oak Ridge National Laboratory. The neutron scattering experiments at the Material and Life Science Experimental Facility, Japan Proton Accelerator Research Complex was performed under a user program (Proposal No. 2019B0011). A portion of computational work was conducted at the Center for Nanophase Materials Sciences which is a DOE Office of Science User Facility. This research used resources of the National Energy Research Scientific Computing Center (NERSC), a U.S. Department of Energy Office of Science User Facility operated under Contract No. DE-AC02-05CH11231. **Author contributions:** S.-H.L.

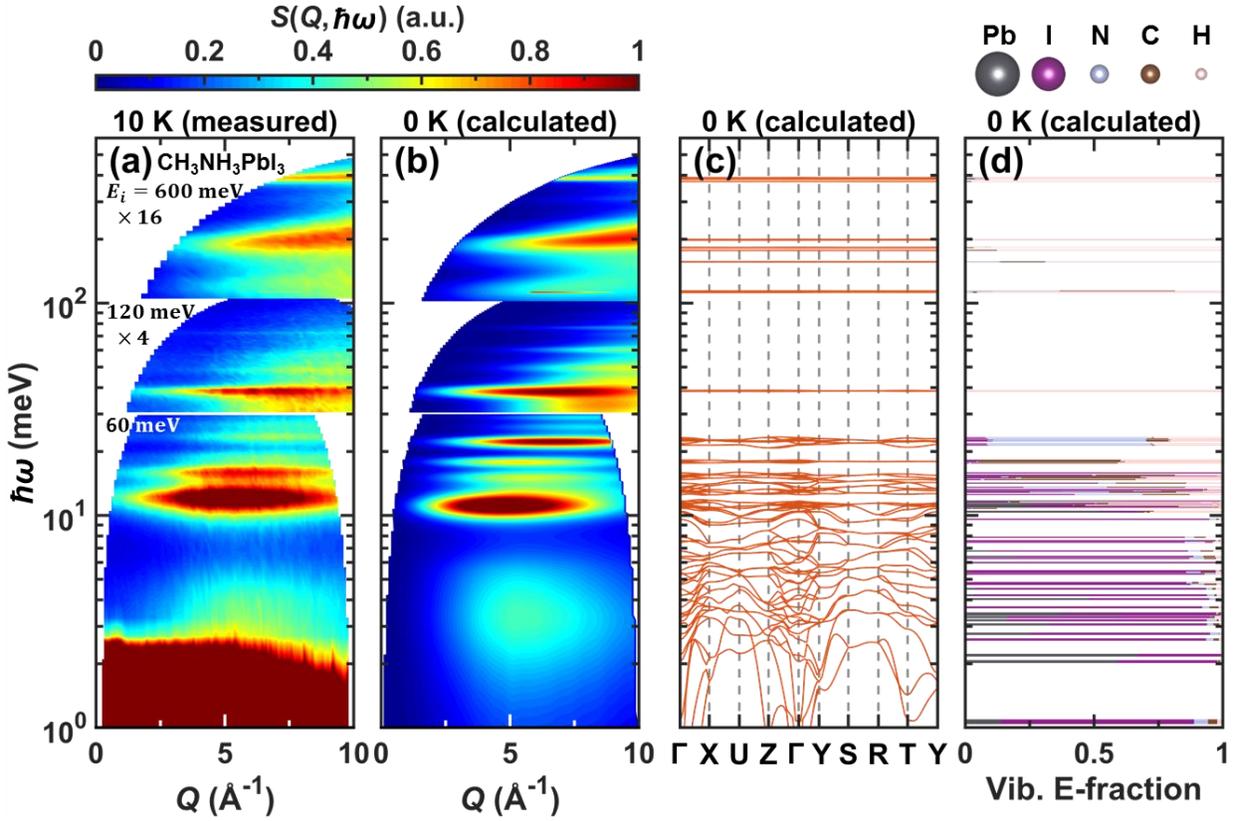

FIG. 1. Phonon spectra of $CH_3NH_3PbI_3$. ((a) and (b)) Color contour maps of the experimental (a) and calculated (b) phonon spectra $S(Q, \hbar\omega)$ of $CH_3NH_3PbI_3$ as a function of the linear momentum transfer $Q$ and the log-scale energy transfer $\hbar\omega$. For better visualization of the entire range of phonon spectra up to ~ 400 meV, $S(Q, \hbar\omega)$ obtained with three different incident neutron energies, $E_i = 60$ meV, 120 meV and 600 meV, are multiplied by different factors, 1, 4 and 16, respectively. (b) DFT phonon spectra calculated for the low-$T$ orthorhombic phase of $CH_3NH_3PbI_3$. (c) DFT phonon band structure of orthorhombic $CH_3NH_3PbI_3$ along high-symmetry $Q$ points. (d) Vibrational energy fractions (Vib. E-fraction) at the $\Gamma$ point for each phonon mode of $CH_3NH_3PbI_3$, (see SI for the calculation details). The width of the gray, violet, cyan, brown and pink bars represent the energy fractions of the Pb, I, N, C and H atoms, respectively.

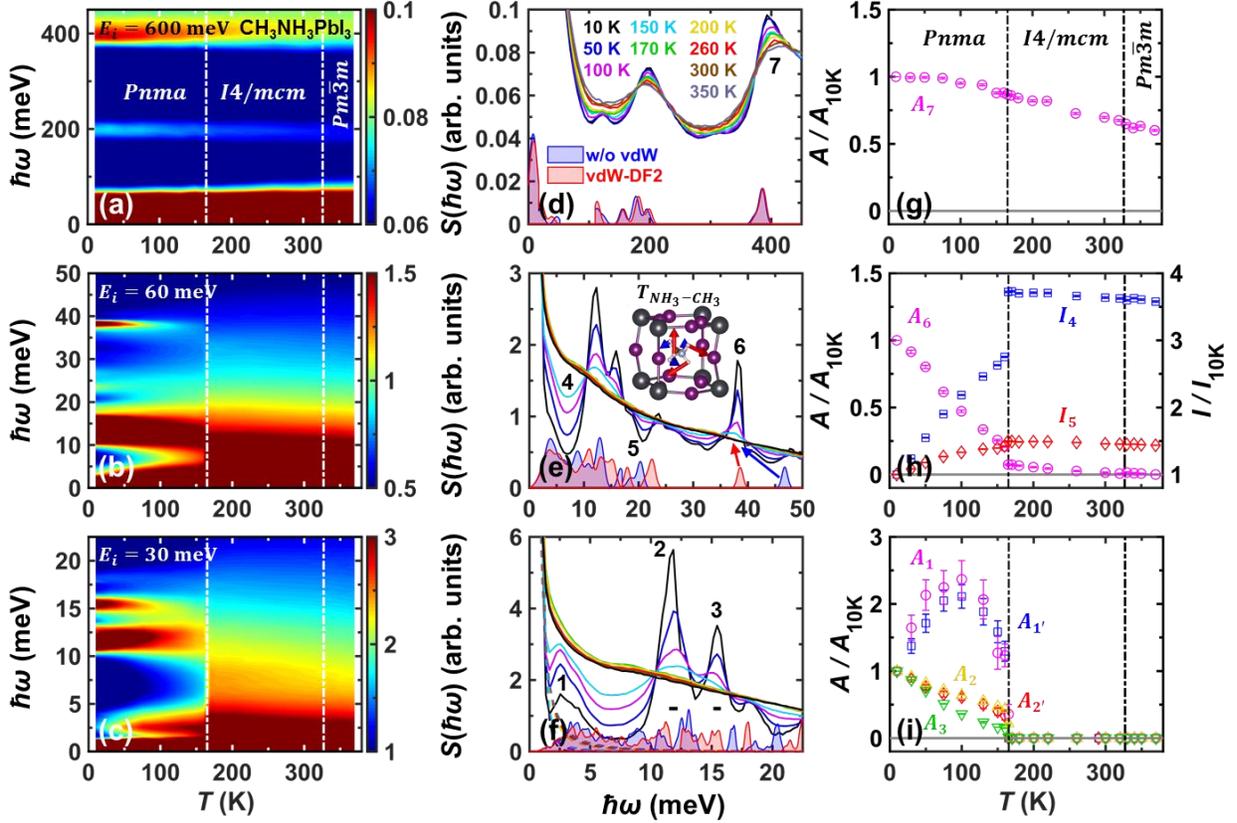

FIG. 2. Temperature-dependent phonon spectra of $CH_3NH_3PbI_3$. ((a) – (c)) Color contour maps of $Q$-integrated neutron scattering intensity, $S(\hbar\omega) = \int S(Q,\hbar\omega)dQ/\int dQ$, as a function of temperature $T$ and $\hbar\omega$, taken at 20 different temperatures while heating from 10 K to 370 K, with the incident neutron energy $E_i$ of 600 meV (a), 60 meV (b) and 30 meV (c). ((d) – (f)) The line plots of the $S(\hbar\omega)$ for nine selected temperatures, 10 K, 50 K, 100 K, 150 K (low-$T$ orthorhombic phase), 170 K, 200 K, 250 K, 300 K (intermediate-$T$ tetragonal phase) and 350 K (high-$T$ cubic phase), taken with $E_i = 600$ meV (d), $E_i = 60$ meV (e) and $E_i = 30$ meV (f). The blue and red lines with shaded areas at the bottom of panels (d) - (f) are the calculated phonon density of states (PDOSs) without vdW correction and with the vdW-DF2 correction [55], respectively. The inset in (e) illustrates the $T_{NH_3-CH_3}$ twisting mode, which corresponds to the phonon peak 6. The dashed lines in (f) represent the contributions from the $CH_3NH_3$ molecule rotations, for all the nine temperatures in the corresponding colors. The black horizontal bars in (f) represent the full-width-at-half-maximum (FWHM) of the instrument resolution for Peak 2 and 3.

((g) – (i)) show the temperature dependence of the fitted Gaussian peak areas, $A$, of phonon mode 1 - 3, 6 and 7, as labeled in (d) – (f). The fitted area $A$ is normalized by the peak area of 10 K, $A_{10K}$. For comparison, the phonon peaks of CD$_3$NH$_3$PbI$_3$ shown in Fig. S5, corresponding to $A_1$ and $A_2$, are shown as the blue squares ($A_{1'}$) and red diamonds ($A_{2'}$) in (i). In (h), the blue squares and red diamonds show the integrated intensity $I$ of the phonon valley 4 ($6 \leq \hbar\omega \leq 8$ meV) and 5 ($20 \leq \hbar\omega \leq 22$ meV) labeled in (e), which are normalized by the values at 10 K, $I_{10K}$. The vertical dash-dotted lines in ((a) – (c)) and ((g) – (i)) represent the transition temperatures for the orthorhombic ($Pnma$)-to-tetragonal ($I4/mcm$) and tetragonal ($I4/mcm$)-to-cubic ($Pm\overline{3}m$) phase transitions, at 165 K and 327 K, respectively.

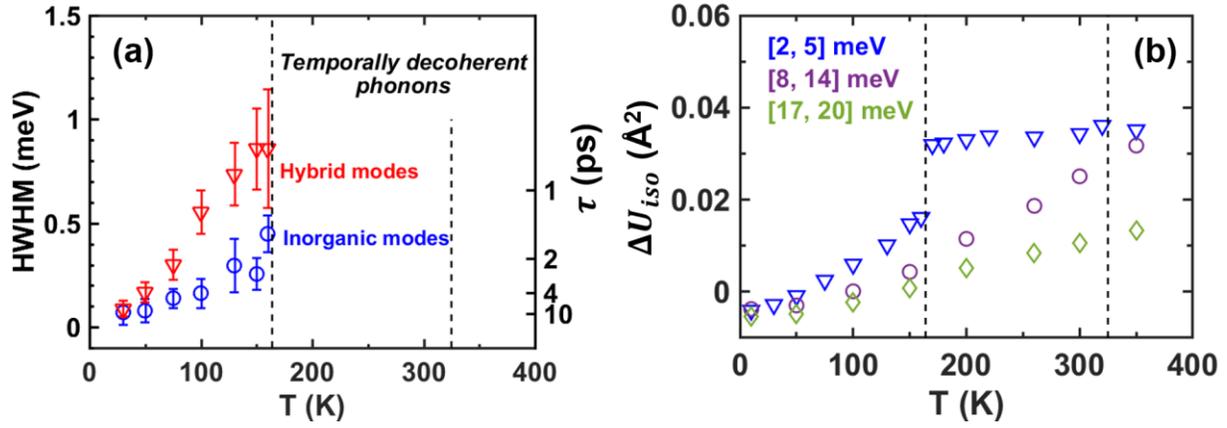

FIG. 3. Temperature-dependent phonon lifetime and atomic displacement parameters of MAPbI$_3$. (a) The blue circles and the red triangles are the phonon energy linewidth and lifetime of the inorganic and hybrid modes, respectively, obtained from fitting $S(\hbar\omega)$ in Fig. S5. (b) The atomic displacement parameters, $\Delta U_{iso} = U_{iso}(T) - U_{iso}(0\ K)$, were obtained from fitting $S(Q)$ in Fig. 4 ((d) – (f)) as described in the text. $\Delta U_{iso}$ are determined and plotted for three different $\hbar\omega$ regions: $2 \leq \hbar\omega \leq 5$ meV ($E_i = 30$ meV, blue triangles), $8 \leq \hbar\omega \leq 14$ meV ($E_i = 60$ meV, violet circles) and $17 \leq \hbar\omega \leq 20$ meV ($E_i = 60$ meV, green diamonds).

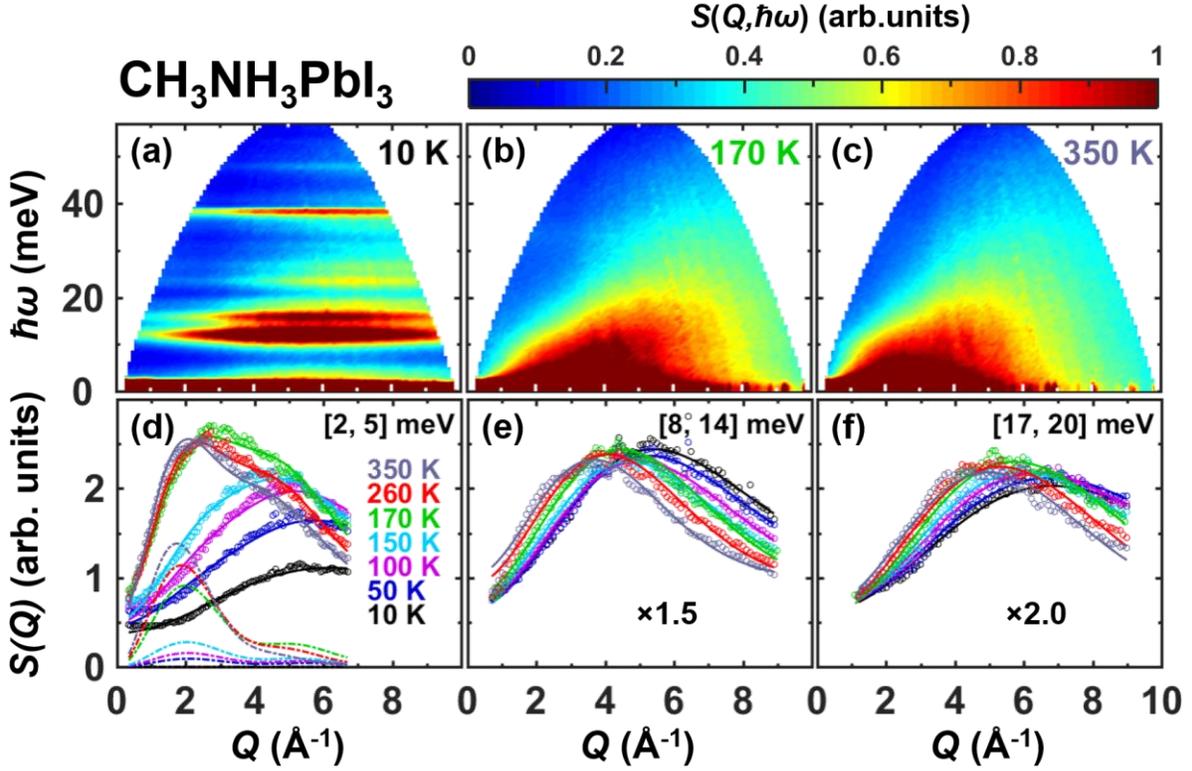

FIG. 4. Inelastic neutron scattering spectra and $Q$-dependence of the phonon spectra for MAPbI$_3$. ((a) – (c)) The color contour maps of $S(Q, \hbar\omega)$ of CH$_3$NH$_3$PbI$_3$ taken at 10 K (orthorhombic phase), 170 K (tetragonal phase), and 350 K (cubic phase) with the incident neutron energy $E_i = 60$ meV. ((d) – (f)) The colored circles are the $\hbar\omega$-integrated neutron scattering intensity, $S(Q)$, for seven different temperatures, taken with the incident neutron energies $E_i = 30$ meV (d) and $E_i = 60$ meV ((e), (f)). The $\hbar\omega$ integration region was $2 \leq \hbar\omega \leq 5$ meV (d), $8 \leq \hbar\omega \leq 14$ meV (e) and $17 \leq \hbar\omega \leq 20$ meV (f), respectively. In (d), the rotational contributions are shown as dash lines and the total fitted intensities are shown as solid lines, as discussed in the text. The panel (e) and (f) are rescaled by a factor of 1.5 and 2.0, respectively, to have the same scale as the panel (d).

# Supplementary Online Material for

# Temporally-decoherent and spatially-coherent vibrations in metal halide perovskite

Depei Zhang, Xiao Hu, Tianran Chen, Douglas L. Abernathy, Ryoichi Kajimoto, Mitsutaka Nakamura, Maiko Kofu, Benjamin J. Foley, Mina Yoon, Joshua J. Choi, and Seung-Hun Lee[*]

*To whom correspondence should be addressed: shlee@virginia.edu

**This PDF file includes:**

- A.  Sample Preparation
- B.  Inelastic Neutron Scattering Experiments
- C.  Group Theoretical Analysis and Rotational Contribution in Neutron Spectra
- D.  Estimation of Phonon Lifetime
- E.  Temperature Dependent Phonon Spectra of Deuterated MAPbI$_3$
- F.  Density-Functional Theory (DFT) Simulations

    F.1. DFT Simulations for Phonon Modes of MAPbI$_3$

    F.2. DFT Calculations of Binding Energies

    F.3. DFT Calculations of Phonon Spectra at Low and High temperatures

Table I - III

Figure S1 - S6

References

# Supporting Online Material

### A. Sample Preparation

For this work, two polycrystalline samples of metal halide perovskite (MHP), an 8 g non-deuterated MAPbI$_3$ (CH$_3$NH$_3$PbI$_3$) and a 0.3 g partially-deuterated MAPbI$_3$ (CD$_3$NH$_3$PbI$_3$), were used in the time-of-flight inelastic neutron scattering (INS) experiments.

The non-deuterated MAPbI$_3$ (CH$_3$NH$_3$PbI$_3$) powder sample was prepared using the same method as reported by Zhu, *et. al.* [72]. The following chemicals were used as received for the CD$_3$NH$_3$PbI$_3$ preparation: hydroiodic acid (57% by weight in water), ethanol, diethyl ether, lead iodide (PbI$_2$) (99.999%) from Sigma Aldrich and methyl-d3-amine hydrochloride (99 atom % D) from MilliporeSigma. CD$_3$NH$_3$I was synthesized through the dissolution of methyl-d3-amine hydrochloride and recrystallization. CD$_3$NH$_3$PbI$_3$ was synthesized via solution crystallization by mixing CD$_3$NH$_3$I with PbI$_2$ in aqueous hydrogen iodide solution and slowly evaporating the liquids via heating in ambient air.

### B. Inelastic Neutron Scattering Experiments

In order to study the vibrational dynamics of CH$_3$NH$_3$PbI$_3$, the inelastic neutron scattering (INS) measurements were performed on both non-deuterated CH$_3$NH$_3$PbI$_3$ and partially deuterated CD$_3$NH$_3$PbI$_3$ samples, using two different time-of-flight neutron scattering instruments.

The INS measurements on CH$_3$NH$_3$PbI$_3$ were performed at the Wide Angular-Range Chopper Spectrometer (ARCS, BL-18) [73] located at the Spallation Neutron Source (SNS) at the Oak Ridge National Laboratory (ORNL). The polycrystalline sample of CH$_3$NH$_3$PbI$_3$ was sealed with helium gas in an annular aluminum can and mounted in a closed cycle refrigerator (CCR). The sample was first cooled to 10 K, then measured at 20 different temperatures upon heating up to 370 K. With the data reduction software package, DAVE Mslice [74], we were able to extract the ARCS INS intensity, $S(Q, \hbar\omega)$, as a function of momentum $Q$, and energy transfers $\hbar\omega$.

The INS measurements on $CD_3NH_3PbI_3$ were performed at the 4D-Space Access Neutron Spectrometer (4SEASONS) [75] located at the Materials and Life Science Experimental Facility (MLF), Japan Proton Accelerator Research Complex (JPARC), with multiple incident energies (MIE) [76].

## C.  Group Theoretical Analysis and Rotational Contribution in Neutron Spectra

The rotation model that accounts for the existence of preferential molecular orientations is called jump model [58, 77]. Since the $MA^+$ molecule is located inside the octahedral cage, its rotational motions are restricted by its own symmetry as well as by the local crystal symmetry of the cage. The possible rotational modes can be accounted for by the irreducible representations of the direct product of the symmetry of the local crystal environment (C) and that of the molecule (M); $\Gamma = C \otimes M$. Here we consider proper rotations only and do not consider improper rotations such as inversion and mirror reflection. For $MAPbI_3$ it was shown [58] that for both cubic and tetragonal phases the simplest jump model that can reproduce the rotational dynamics is $\Gamma = C \otimes M = C_4 \otimes C_3$ where $C_4$ and $C_3$ represent the four-fold symmetry of the cuboctahedral cage and the three-fold symmetry of the molecule, respectively. Since this point group theory has been extensively described in many textbooks, we will state here only the basic formalism that is necessary for our discussion. In the group theory, the static and dynamic structure factor for rotational motions of molecules embedded in a crystal can be written as [58, 77]

$$S(Q, \hbar\omega) = e^{-\langle u^2 \rangle Q^2} \left( \sum_\gamma A_\gamma(Q) \frac{1}{\pi} \frac{\tau_\gamma}{1+\omega^2 \tau_\gamma^2} \right) \qquad (S1)$$

where the sum over $\gamma$ runs over all the irreducible representations of $\Gamma$, $\Gamma_\gamma$. For a polycrystalline sample, $A_\gamma(Q)$ is given by [77]

$$A_\gamma(Q) = \frac{l_\gamma}{g}\sum_\alpha \sum_\beta \chi_\gamma^{\alpha\beta} \sum_{C_\alpha} \sum_{M_\beta} j_0(Q|\mathbf{R} - C_\alpha M_\beta \mathbf{R}|) \qquad (S2)$$

Here $g$ is the order of $\Gamma$ and $l_\gamma$ is the dimensionality of $\Gamma_\gamma$. The sums over $\alpha$ and $\beta$ run over all the classes of C and M, respectively, and the sums over $C_\alpha$ and $M_\beta$ run over all the rotations that belong to the class of the crystal symmetry group, $\alpha$, and to the class of the molecule symmetry group, $\beta$, respectively. The characters of $\Gamma_\gamma$, $\chi_\gamma^{\alpha\beta}$, are the products of the characters of $C_{\gamma C}$ and $M_{\gamma M}$, $\chi_{\gamma C}^\alpha$ and $\chi_{\gamma M}^\beta$, respectively; $\chi_\gamma^{\alpha\beta} = \chi_{\gamma C}^\alpha \chi_{\gamma M}^\beta$. $j_0(x)$ is the zeroth spherical Bessel function and, $|\mathbf{R} - C_\alpha M_\beta \mathbf{R}|$, called the jump distance, is the distance between the initial atom position $\mathbf{R}$, and final atom position $C_\alpha M_\beta \mathbf{R}$. The relaxation time for the $\Gamma_\gamma$ mode, $\tau_\gamma$, can be written in terms of the relaxation times for $C_\alpha$ and $M_\beta$, $\tau_\alpha$ and $\tau_\beta$, respectively, [77]

$$\frac{1}{\tau_\gamma} = \sum_\alpha \frac{n_\alpha}{\tau_\alpha}\left(1 - \frac{\chi_\gamma^{\alpha e}}{\chi_\gamma^{Ee}}\right) + \sum_\beta \frac{n_\beta}{\tau_\beta}\left(1 - \frac{\chi_\gamma^{E\beta}}{\chi_\gamma^{Ee}}\right) \qquad (S3)$$

where $n_\alpha$ and $n_\beta$ are the number of symmetry rotations that belong to the classes, $\alpha$ and $\beta$, respectively. $E$ and $e$ represent the identity operations of C and M, respectively.

Note that once the relaxation times, $\tau_\gamma$, are determined, then $S(Q, \hbar\omega)$ is determined for all values of $Q$ and $\hbar\omega$, besides the Debye-Waller factor $e^{-\langle u^2 \rangle Q^2}$ that is to be fitted for each temperature. And the relaxation times, $\tau_\gamma$, can be determined by fitting the elastic data, $S(Q, \hbar\omega = 0)$, to $S(Q, \hbar\omega = 0) = e^{-\langle u^2 \rangle Q^2}\left(\sum_\gamma A_\gamma(Q)\frac{\tau_\gamma}{\pi}\right)$. Fig. S1 shows the results of the fitting for four different temperatures, and the fitted $\tau_\gamma$'s for all the temperatures taken are listed in Table I. And $S(Q, \hbar\omega)$ for non-zero $\hbar\omega$ can be obtained by Eq. S1. Fig. S2 and S3 show that the theoretical $S(Q, \hbar\omega)$ convoluted with instrumental energy resolution (colored surface) reproduces the experimental data (black dots) for all temperatures. The color-coded dashed lines in Fig. 2 (f) and Fig. S5 are the $Q$-integrated theoretical $S(Q, \hbar\omega)$ convoluted with the instrumental energy resolution.

Table I. Estimated relaxation times, $\tau_{C_4}$ and $\tau_{C_3}$, and the square-root of the mean-squared displacement $\sqrt{\langle u^2 \rangle}$ for the rotational dynamics of $CD_3NH_3^+$ and $CH_3NH_3^+$ cations that are extracted from fitting the elastic channel data as discussed in the text. Values in the parentheses indicate their errors.

|  | $CD_3NH_3^+$ | | | $CH_3NH_3^+$ | | |
|---|---|---|---|---|---|---|
| $T$ (K) | $\tau_{C_3}$ (ps) | $\tau_{C_4}$ (ps) | $\sqrt{\langle u^2 \rangle}$ (Å) | $\tau_{C_3}$ (ps) | $\tau_{C_4}$ (ps) | $\sqrt{\langle u^2 \rangle}$ (Å) |
| 350 | 6.22(6) | 6.34(6) | 0.396(1) | 1.97(1) | 7.32(8) | 0.374(1) |
| 300 | 7.16(8) | 9.2(1) | 0.373(1) | 2.44(1) | 9.4(1) | 0.361(1) |
| 260 | 7.75(9) | 13.7(3) | 0.349(1) | 2.75(1) | 14.6(3) | 0.346(1) |
| 200 | 9.4(1) | 28(1) | 0.321(2) | 3.43(2) | 24(1) | 0.332(2) |
| 150 | 31(2) | ∞ | 0.225(2) | 11.8(2) | ∞ | 0.261(2) |
| 100 | 110(18) | ∞ | 0.173(3) | 22(1) | ∞ | 0.215(2) |
| 50 | ∞ | ∞ | 0.118(4) | 39(2) | ∞ | 0.173(3) |
| 10 | ∞ | ∞ | 0.075(6) | ∞ | ∞ | 0.145(3) |

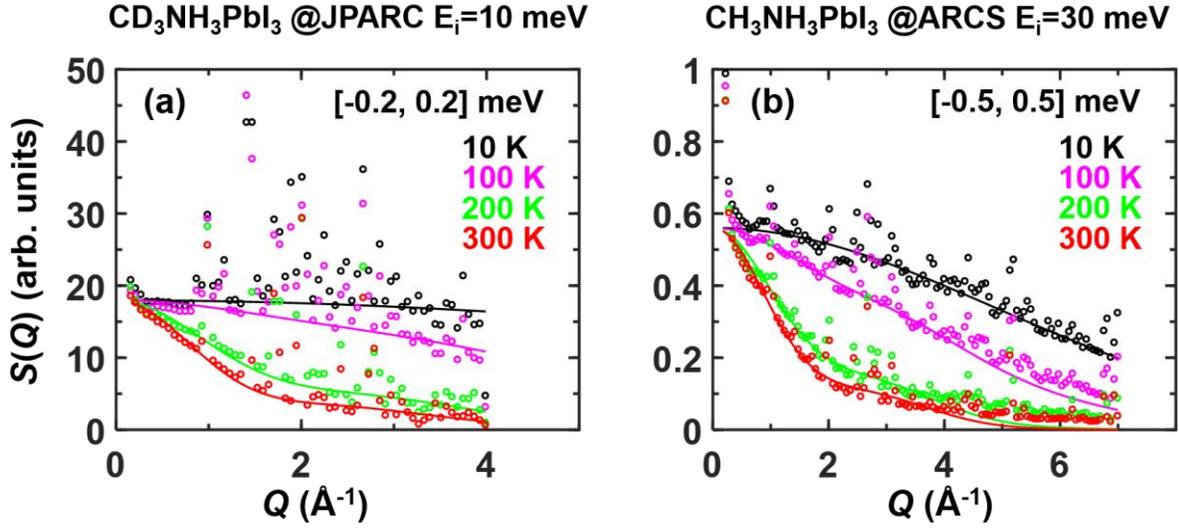

FIG. S1. Temperature-dependent elastic channel data. The colored circles showed $\hbar\omega$-integrated neutron scattering intensity, $S(Q)$, at 4 selected temperatures, 10 K, 100 K, 200 K and 300 K with $-0.2 \leq \hbar\omega \leq 0.2$ meV for $CD_3NH_3PbI_3$ data taken at JPARC (a); $-0.5 \leq \hbar\omega \leq 0.5$ meV for $CH_3NH_3PbI_3$ taken at ARCS (b). The colored solid lines are the rotational contributions at the corresponding temperatures, obtained from the fitting to the jump model described in the text.

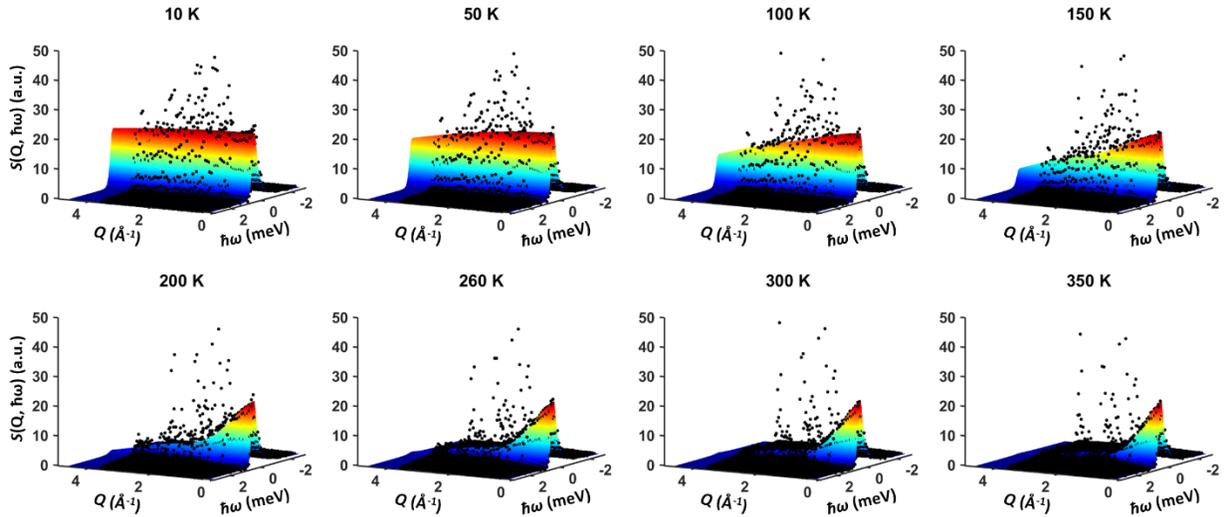

FIG. S2. Inelastic neutron scattering intensity and rotational contributions for $CD_3NH_3PbI_3$. The data (black dots) were collected at 10 K, 50 K, 100 K, 150 K (orthorhombic phase $Pnma$), 200 K, 260 K, 300 K (tetragonal phase $I4/mcm$), and 350 K (cubic phase $Pm\bar{3}m$) with the incident neutron energy $E_i = 10$ meV at JPARC. The colored surface is the fitted rotational contribution based on the fitted parameters listed in Table I.

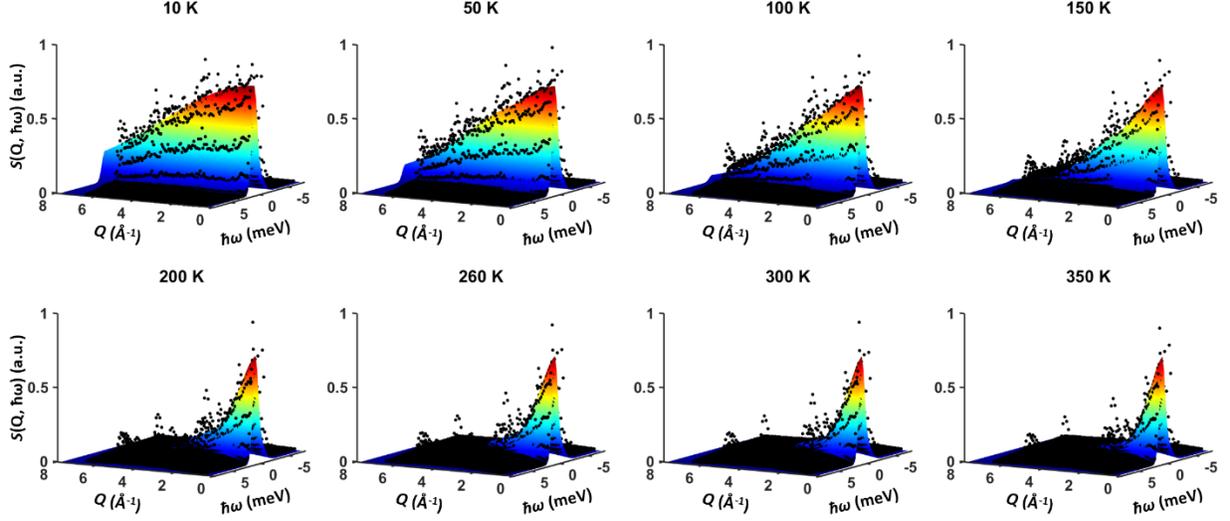

FIG. S3. Inelastic neutron scattering intensity and rotational contributions for $CH_3NH_3PbI_3$. The data (black dots) were collected at 10 K, 50 K, 100 K, 150 K (orthorhombic phase $Pnma$), 200 K, 260 K, 300 K (tetragonal phase $I4/mcm$), and 350 K (cubic phase $Pm\bar{3}m$) with the incident neutron energy $E_i = 30$ meV at ARCS. The colored surface is the fitted rotational contribution based on the fitted parameters listed in Table I.

### D. Estimation of Phonon Lifetime

To estimate the phonon lifetime, we used the neutron scattering cross section $S(Q, \hbar\omega)$ taken for $CD_3NH_3PbI_3$ with two different energies of incident neutrons, $E_i = 10$ meV and 30 meV. In order to evaluate the well-defined phonon peaks (such as intensity and lifetime), we fitted the two $E_i = 10$ meV and 30 meV data sets separately using Voigt function $V(x; \sigma, \gamma)$ for each phonon peak. A Voigt function is the convolution of a Gaussian function and a Lorentzian function. Now, since our main interest is how the temporal coherence evolves with temperature and at 10 K the phonons have perfect temporal coherence, we fitted the 10 K phonon spectra as Gaussian functions and took that as the powder-averaged energy spectra of phonons with perfect coherence convoluted with instrumental energy resolutions. And then, the temperature-induced broadening of the phonon peaks at higher temperatures was fitted with Lorentzian functions at higher temperatures. To minimize the number of fitting parameters and thus to increase the reliability of the analysis, we introduce two lifetimes only to account for the optical inorganic and hybrid modes, assuming that all the inorganic modes have the same lifetime and all the hybrid modes have another same lifetime. This is a reasonable assumption because the inorganic modes involve Pb-I-Pb bending and weaker vibrations of the organic molecule while the hybrid modes involve Pb-I-Pb stretching with stronger vibrations of the molecule. The Half-Width-at-Half-Maximum (HWHM) of the Lorentzian becomes the inverse of the phonon lifetime; $\tau = \frac{\hbar}{HWHM}$.

The Voigt function $V(x; \sigma, \gamma)$ can be written in terms of the Faddeeva function, $w(z)$:

$$V(x; \sigma, \gamma) = 2\sqrt{(\ln 2)} \frac{Re[w(z)]}{\sigma \sqrt{\pi}} \quad (S4)$$

where $\sigma$ is the FWHM (full-width-at-half-maximum) of the Gaussian function, $G(x) = \frac{2\sqrt{(\ln 2)}}{\sigma \sqrt{\pi}} e^{-\frac{4(\ln 2)x^2}{\sigma^2}}$, and $\gamma$ is the HWHM (half-width-at-half-maximum) of the Lorentzian function, $L(x) = \frac{1}{\pi} \frac{\gamma}{x^2 + \gamma^2}$. In the formula above, $z$ is evaluated as $z = 2\sqrt{(\ln 2)} \left(\frac{x+i\gamma}{\sigma}\right)$, and the Faddeeva function $w(z) = e^{-z^2} \left(1 + \frac{2i}{\sqrt{\pi}} \int_0^z e^{t^2} dt\right)$. The fitting results for the two datasets are plotted for several different temperatures in Fig. S4. The HWHM (half-width-at-half-maximum) of the fitted Lorentzian functions and corresponding calculated phonon lifetime of two selected phonon peaks are listed in Table II.

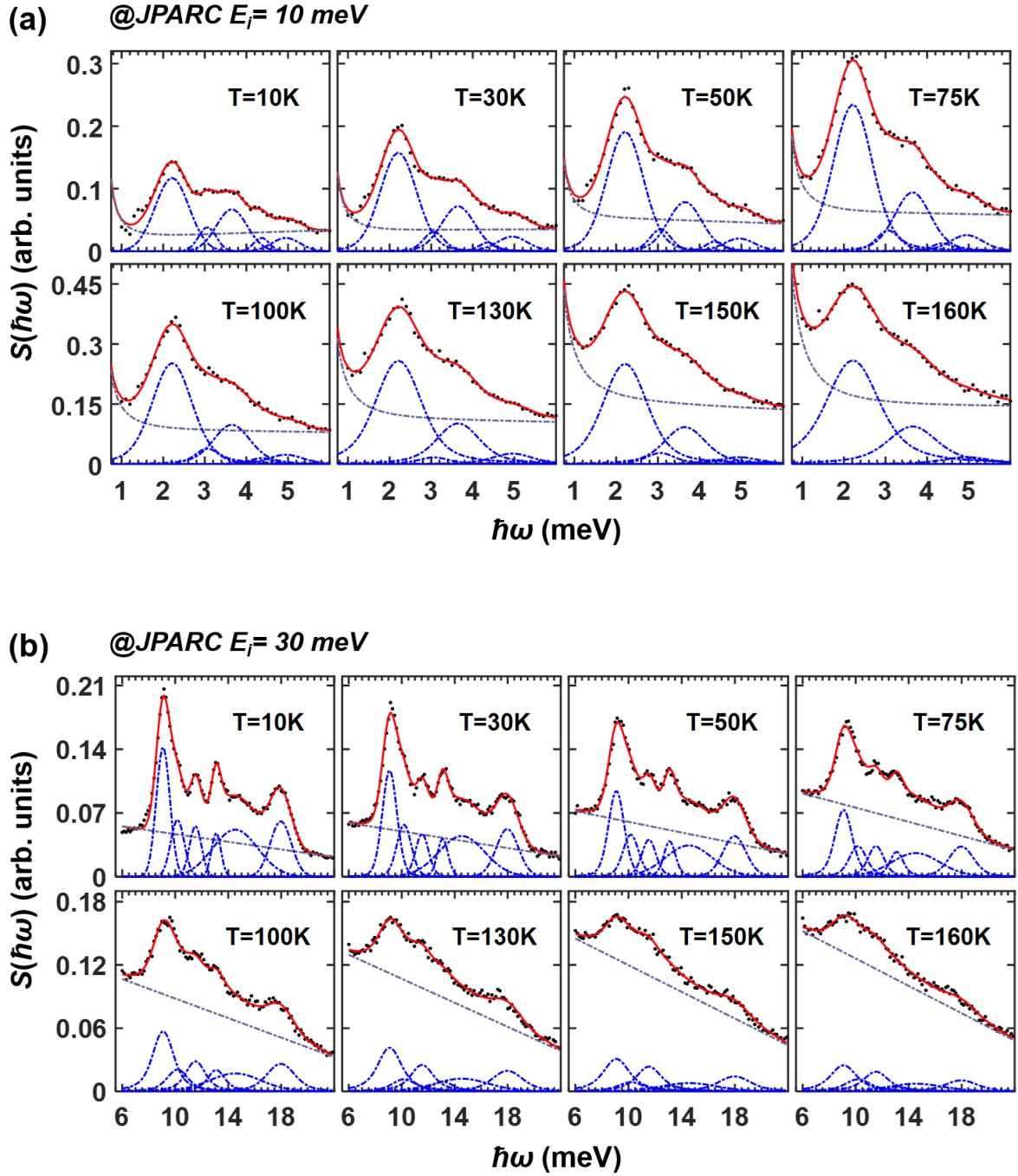

FIG. S4. Phonon spectra fittings of $CD_3NH_3PbI_3$. The momentum-integrated experimental phonon spectra (black dots), $S(\hbar\omega)$, were collected at the time-of-flight spectrometer 4SEASONS at JPARC with two incident neutron energies $E_i = 10$ meV and $E_i = 30$ meV. (a) For the $E_i = 10$ meV dataset, the phonon spectra were fitted with 5 Voigt functions (blue dashed lines on the bottom). The grey dashed lines account for the increasing rotational contributions and liquid-like continuum with increasing temperature. (b) For the $E_i = 30$ meV dataset, the rotational

contribution in the fitted energy range is negligible. The phonon spectra were fitted with 6 Voigt functions (blue dashed lines). The grey dashed lines account for the liquid-like continuum with increasing temperature. In (a) and (b), the red solid lines are the fitted total intensities.

Table II. Estimated phonon lifetimes of the inorganic and hybrid phonon modes in the orthorhombic phase of MAPbI$_3$, extracted from the phonon model fittings on data taken at JPARC. Values in the parentheses indicate their errors.

| T (K) | Inorganic modes | | Hybrid modes | |
|---|---|---|---|---|
| | Lorentzian HWHM (meV) | Phonon lifetime (ps) | Lorentzian HWHM (meV) | Phonon lifetime (ps) |
| 160 | 0.45(9) | 1.47 | 0.86(28) | 0.77 |
| 150 | 0.26(8) | 2.58 | 0.86(20) | 0.77 |
| 130 | 0.30(13) | 2.22 | 0.74(15) | 0.90 |
| 100 | 0.16(7) | 4.07 | 0.55(10) | 1.19 |
| 75 | 0.14(5) | 4.76 | 0.30(7) | 2.20 |
| 50 | 0.08(6) | 8.19 | 0.17(5) | 3.90 |
| 30 | 0.07(6) | 9.57 | 0.08(4) | 7.50 |

### E. Temperature Dependent Phonon Spectra of Deuterated MAPbI$_3$

Similar as the temperature-dependent phonon spectra measurement of the non-deuterated MAPbI$_3$ (CH$_3$NH$_3$PbI$_3$), which is shown in Fig. 2 of the main text, we have also performed such experiment on a 0.3 g powder sample of partially deuterated MAPbI$_3$ (CD$_3$NH$_3$PbI$_3$), that, due to the deuteration, would yield weaker incoherent rotational signals. Fig. S5 shows momentum-integrated $S(T,\hbar\omega)$, obtained from two data sets taken with $E_i =$ 10 meV and 30 meV, at various $T$s, to focus on how the inorganic and hybrid modes evolve with increasing $T$. Compared to the case of CH$_3$NH$_3$PbI$_3$, those modes of CD$_3$NH$_3$PbI$_3$, notably the hybrid modes, shift to lower energies due to the replacement of H by heavier D. The color-coded dashed lines are the rotational contribution determined by the point group theory analysis (for details, see Session C). Upon warming, the integrated intensities of the inorganic modes and hybrid modes, $A'_1$ and $A'_2$, respectively, show $T$-dependences identical to those of non-deuterated CH$_3$NH$_3$PbI$_3$, as shown in Fig. 2 (i) in the main text.

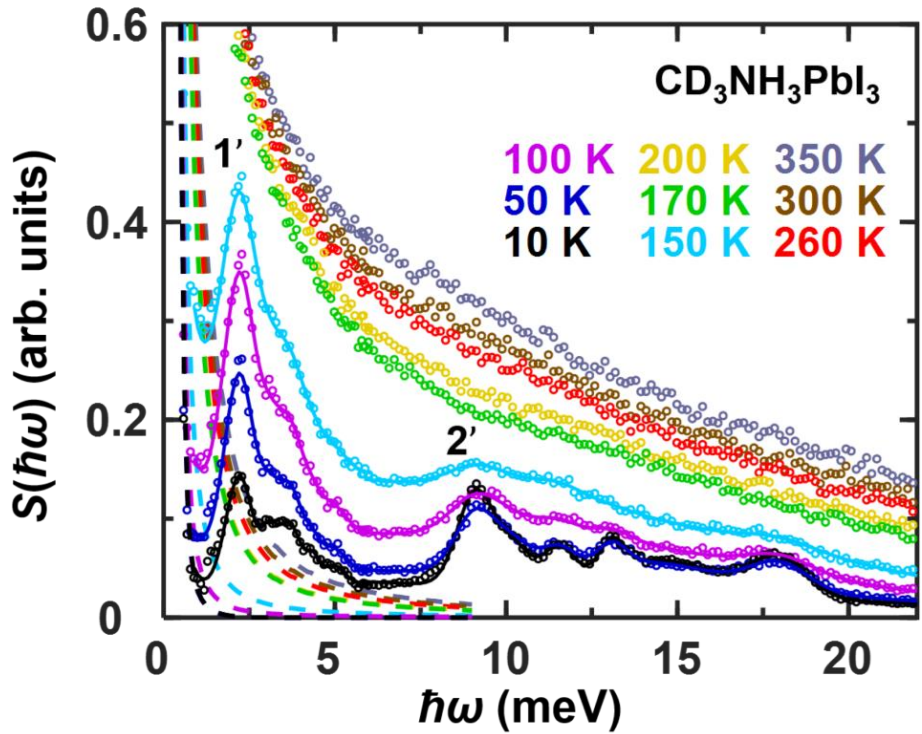

FIG. S5. Temperature-dependent phonon spectra of CD$_3$NH$_3$PbI$_3$. The colored circles are the $Q$-integrated scattering cross section, $S(\hbar\omega)$, obtained from the powder sample of CD$_3$NH$_3$PbI$_3$ with two different incident neutron energies $E_i = 10$ meV and $E_i = 30$ meV, for nine selected

temperatures, 10 K, 50 K, 100 K, 150 K (orthorhombic phase), 170 K, 200 K, 260 K, 300 K (tetragonal phase) and 350 K (cubic phase). To connect the two datasets in $\hbar\omega$-space, we compared the two $S(\hbar\omega)$ for each $T$ over their overlapping energy region around $\hbar\omega \sim 6\ meV$ and obtained a scaling factor to normalize the two data sets in the same unit. The colored dashed lines are the rotation contributions (for details, see Session C). The colored solid lines are the results of fitting described in the text.

Note that upon warming, the phonon peaks, notably the hybrid modes between 7.5 meV and 20 meV, broaden before they become the energy continuum above 170 K in the tetragonal and the cubic phase. This means the temporal-decoherence of those phonons occurs gradually in the orthorhombic phase. In order to quantify the gradual temporal-decoherence, we fitted $S(\hbar\omega)$ to a sum of several Voigt functions each of which is the convolution of a Gaussian function and a Lorentzian function. Here, the Gaussian function accounts for the powder-averaged signal from phonons with perfect temporal-coherence convoluted with the instrumental energy resolution. Since at 10 K the phonons are temporally coherent as evidenced by the well-defined peaks in $\hbar\omega$-space, the Gaussian functions in the Voigt functions were determined by fitting the profile at 10 K, $S(10\ K, \hbar\omega)$, to simple Gaussians. The Lorentzian functions at higher temperatures, then, represent the $T$-induced broadening of the phonon peaks, and the Half-Width-at-Half-Maximum (HWHM) is the inverse of the phonon lifetime. See Session D for detailed description of the fitting. The fitting results are shown as the color-coded solid lines in Fig. S5. As plotted in Fig. 3 (a) in the main text, the optical phonon lifetimes decrease gradually upon warming in the orthorhombic phase before the optical phonons lose their temporal coherence in the tetragonal phase. Note that the lifetime of the hybrid modes that involve stronger vibrations of the organic molecules decreases faster than that of the optical mostly-inorganic modes.

## F. Density-Functional Theory (DFT) Simulations

### F.1. DFT Simulations for Phonon Modes of MAPbI$_3$

In order to understand the atomic nature of the vibrational modes and the experimental INS spectra of non-deuterated MAPbI$_3$ (CH$_3$NH$_3$PbI$_3$), first-principles density functional theory (DFT) calculations were performed. The DFT calculations were based on the Vienna *ab initio* Simulation Package (VASP) [78] with projector augmented wave method [79]; generalized gradient approximation in the form of Perdew-Burke-Ernzerhof (PBE) is adopted for the exchange-correlation functional [80], the energy cutoff of the plane-wave basis set was 400 eV, and a ***k***-mesh with ***k***-spacing less than 0.2 Å$^{-1}$ was used for the self-consistent calculations. After the self-consistent calculations, the phonon eigenmodes were then analyzed with the finite displacement method [81, 82] as implemented in Phonopy [83], an open-source package. And we found that the van der Waals (vdW) interaction plays a significant role in capturing the experimental INS spectra. A specific vdW density functional, vdW-DF2 [52-55], was employed to account accurately the non-local vdW corrections.

In the DFT calculations, the experimentally refined low-temperature structures were used as the initial structure; the orthorhombic (*Pnma*) structure at 4 K for CH$_3$NH$_3$PbI$_3$ with $(a, b, c) = (8.816, 12.598, 8.564)$ Å [58]. The initial atomic configurations were fully relaxed within the given space group symmetry, *Pmna*. A universal scaling factor $S_{LC} = 0.99$, was applied to the lattice constants $(a, b, c)$, which reproduced best the experimental phonon spectra. The phonon densities of states (PDOSs) were then calculated via Phonopy [83], using an $8 \times 8 \times 8$ grid of the phonon momentum ***q*** space. The phonon eigenvalues and eigenvectors from the Phonopy output were then transferred to a third-party package, OCLIMAX [59], for the phonon spectrum simulations. Multiple-phonon scatterings, up to 10 phonon excitations, are considered during the OCLIMAX simulations.

All the DFT calculation results with the vdW-DF2 correction for MAPbI$_3$, including the phonon density of states (PDOS), and the atomic vibration energy fractions and the animations of the 144 eigenmodes, are shown on S.-H. Lee's website [56]. Below, we describe some details of the calculation methods.

The atomic vibrational energy fractions (Vib. E-fraction) for MAPbI$_3$ that represent the contributions of different atoms to the energy of each phonon mode is defined as the following,

$$\text{VEF}_a(s, \boldsymbol{q}) = \frac{\sum_{i \in t} m_i \omega_s(\boldsymbol{q})^2 |\boldsymbol{u}_i(s,\boldsymbol{q})|^2}{\sum_{i \in all} m_i \omega_s(\boldsymbol{q})^2 |\boldsymbol{u}_i(s,\boldsymbol{q})|^2} = \frac{\sum_{i \in t} |\boldsymbol{e}_i(s,\boldsymbol{q})|^2}{\sum_{i \in all} |\boldsymbol{e}_i(s,\boldsymbol{q})|^2}. \tag{S5}$$

Here, the subscript $a$ represents each atomic type ($a = $ Pb, I, N, C, H), $s$ and $\boldsymbol{q}$ represent the phonon mode index and phonon wavevector, respectively. $\boldsymbol{u}_i(s,\boldsymbol{q})$ is the displacement of the $i$-th atom due to the activation of phonon mode $s$. At $T \to 0$, $\boldsymbol{u}_i(s,\boldsymbol{q}) = \sqrt{\frac{\hbar}{2 m_i \omega_s(\boldsymbol{q})}} \boldsymbol{e}_i(s,\boldsymbol{q})$ [83-85], where $\hbar$ is the reduced Planck constant, $m_i$ is the mass of the $i$-th atom, $\omega_s(\boldsymbol{q})$ is the eigen frequency of phonon mode $s$ at $\boldsymbol{q}$, and $\boldsymbol{e}_i(s,\boldsymbol{q})$ is the phonon polarization vector. The atomic vibrational energy fractions were computed at the $\Gamma$-point with the vdW-DF2 correction and are shown in Fig. 1 (d) of the main text.

In addition, to categorize the phonon modes we have also decomposed the vibrational energies into different types of vibrations, similarly to the method introduced in Ref. [48]; Pb-I-Pb translation, Pb-I-Pb rocking, Pb-I-Pb bending, Pb-I-Pb stretching and MA translation, MA spinning, MA libration, MA internal modes. The animations for all the different types of vibrations are shown on S.-H. Lee's website [56].

### F.2. DFT Calculations of Binding Energies

The binding energies of the chemical bond between atom $X$ and $Y$ can be calculated as

$$E_b(X - Y) = E(X) + E(Y) - E(XY), \tag{S6}$$

where $E(X)$, $E(Y)$ and $E(XY)$ are the total electronic energy of single $X$, $Y$ atoms and $X - Y$ pair, which can be obtained from DFT calculations. We calculated the binding energies of the Pb-I and I-H bonds in MAPbI$_3$, and the ionic Na-Cl bonds in the ionic NaCl lattice, as well as the covalent Si-O bonds in the covalent SiO$_2$ lattice. The crystal structures of NaCl and SiO$_2$ (SiO$_2$-HP) used in our calculations were obtained from the Phonopy examples [86].

As is shown in Table III, the binding energies of Pb-I and I-H bonds are ~ 2.99 and 3.20 eV for MAPbI$_3$, which is close to the 3.95 eV Na-Cl ionic bond. The covalent Si-O bond, on the other hand, has a binding energy of 9.95 eV, which is more than 2 times stronger than the Pb-I and I-H bonds. Therefore, we conclude that the Pb-I and I-H bonds in MAPbI$_3$ are weak ionic bonds. On the other hand, the intramolecular bonds in MAPbI$_3$ have similar binding energies as the Si-O bonds in the SiO$_2$ lattice, indicating that the intramolecular bonds are covalent.

Table III. Chemical bonds in MAPbI$_3$, NaCl and SiO$_2$. The chemical bond information for MAPbI$_3$, NaCl and SiO$_2$ are listed below. The bond lengths between atom pairs $X - Y$ are extracted from the crystal structures for these systems.

| System | Chemical Bond | | | |
| --- | --- | --- | --- | --- |
|  | Atom Pair | Length (Å) | Energy (eV) | Bond Type |
| NaCl | Na-Cl | 2.85 | 3.95 | Ionic |
| MAPbI$_3$ | Pb-I | 3.25 | 2.99 |  |
|  | I-H | 2.63 | 3.20 |  |
|  | N-C | 1.52 | 9.03 | Covalent |
|  | C-C | 1.54 | 7.12 |  |
|  | N-H | 1.04 | 5.91 |  |
|  | C-H | 1.10 | 5.35 |  |
| SiO$_2$ | Si-O | 1.73 | 9.95 |  |

### F.3. DFT Calculations of Phonon Spectra at Low and High temperatures

To investigate the phonon melting behaviors, we simulated the temperature dependence of the powder averaged phonon spectra of MAPbI$_3$ using OCLIMAX [59]. For comparison purposes, similar temperature dependent phonon spectra were calculated for the ionic NaCl lattice and the covalent $SiO_2$ lattice.

Fig. S6 shows the simulated temperature dependent phonon spectra for $MAPbI_3$, NaCl and $SiO_2$, respectively. As plotted in Fig. S6 (a), (e), (i) and (b), (f), (j), the calculated phonon spectra at 0 K and 300 K for the three systems show clear similarities between $MAPbI_3$ and NaCl, the phonon intensities reduced prominently at 300 K. For the covalent $SiO_2$ lattice, the difference between phonon spectra at 0 K and 300 K is much smaller. The momentum transfer ($Q$) integrated phonon spectra, $S(\hbar\omega) = \int_0^{15 \text{ Å}^{-1}} S(Q, \hbar\omega) dQ$ are shown in Fig. S6 (c), (g) and (k) for $MAPbI_3$, NaCl and $SiO_2$, respectively, where the blue and red solid lines represent total phonon spectra, $S(\hbar\omega)$, at 0 K and 300 K, the dash-dotted lines are the corresponding single phonon contributions. In order to quantitatively analyze the temperature dependence, we performed the Gaussian peak fittings for three different single phonon peaks, as labeled as 1 - 3 in Fig. S6 (c), (g) and (k) for the three systems. The fitted peak areas $A$ are shown in Fig. S6 (d), (h) and (l). For $MAPbI_3$ and NaCl, the peak areas, $A$s, show a gradual decrease as temperature increases, while for $SiO_2$, $A$s exhibit a tiny decrease ($A_2$ and $A_3$) and even a nearly linear increase ($A_1$) up to 500 K. It should be noted the raise of $A_1$ for $SiO_2$ (and for NaCl, below 100 K) can be explained by the increasing phonon population at high temperatures. Our temperature dependent phonon spectra simulations show that the gradually decreasing phonon intensities, i.e., the phonon melting behavior, is a general feature for the ionic lattices with the weak ionic bonds. The phonon modes, in the covalent lattices, e.g., $SiO_2$, on the other hand, can survive at high temperatures.

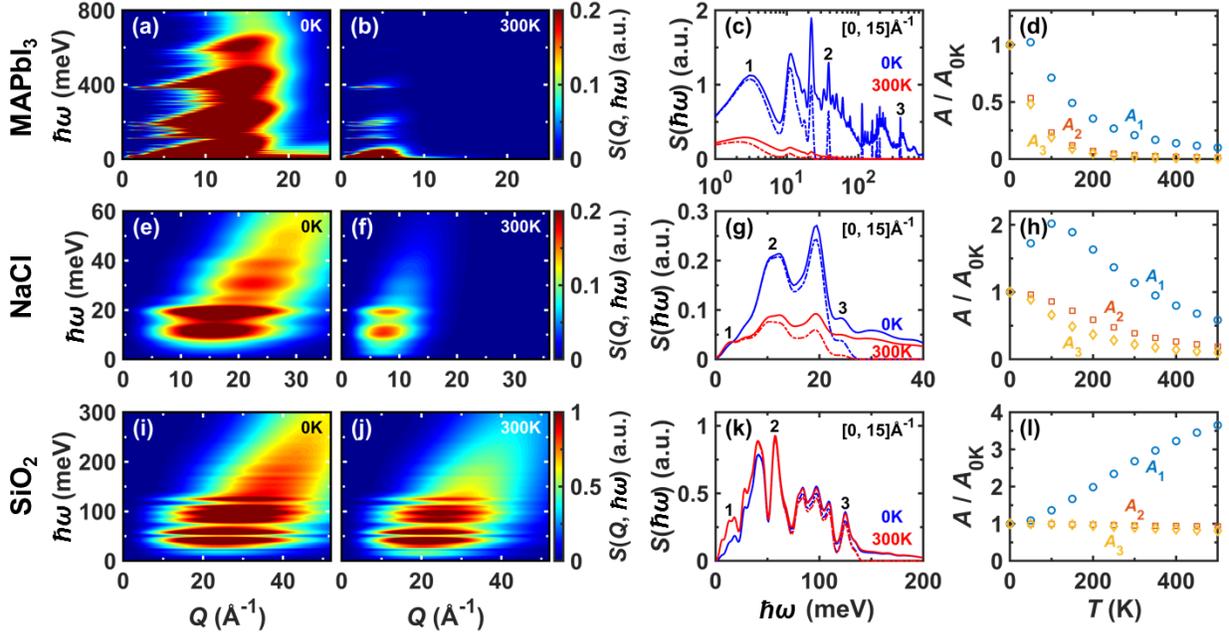

FIG. S6. Simulated phonon spectra for MAPbI$_3$, NaCl and SiO$_2$. ((a), (e), (i)) and ((b), (f), (j)) Contour maps of the simulated inelastic neutron spectra, $S(Q, \hbar\omega)$, for MAPbI$_3$, NaCl, SiO$_2$ at 0 K and 300 K, respectively, which are calculated with OCLIMAX [59]. ((c), (g), (k)) The momentum transfer $Q$-integrated phonon spectra, $S(\hbar\omega) = \int_0^{15\,\text{Å}^{-1}} S(Q, \hbar\omega) dQ$, for MAPbI$_3$, NaCl, SiO$_2$. The blue and red solid lines in (c), (g), (k) represent total phonon spectra, $S(\hbar\omega)$, at 0 K and 300 K, while the dash-dotted lines are the corresponding single phonon contributions. ((d), (h), (l)) The fitted Gaussian peak areas $A$ of the phonon peaks (1) - (3), which are labeled in (c), (g), (k) for MAPbI$_3$, NaCl, SiO$_2$, respectively. The fitted areas $A$ are normalized by the peak area at 0 K, $A_{0\text{K}}$.